\documentclass{edm_article}

\usepackage{booktabs}
\usepackage{subcaption}
\usepackage{url}
\begin{document}

\title{Multi-Agent Collaborative Framework For Math Problem Generation}

\numberofauthors{4}
\author{
    Kia Karbasi\\
    \affaddr{University of California, Los Angeles}\\
    \email{kiakarbasi@ucla.edu}
    \and
    Kevin Hong\\
    \affaddr{University of California, Los Angeles}\\
    \email{kevinhong1167@ucla.edu}
    \and
    Mohammad Amin Samadi\\
    \affaddr{University of California, Irvine}\\
    \email{masamadi@uci.edu}
    \and
    Gregory Pottie\\
    \affaddr{University of California, Los Angeles}\\
    \email{pottie@ee.ucla.edu}
}

\maketitle

\begin{abstract}
Automatic question generation (AQG) for mathematics education remains an elusive goal for Intelligent Tutoring Systems and educators. While pre-trained transformer-based language models have significantly advanced natural language generation, they often struggle to precisely control problem complexity and cognitive demands.  In this paper, we introduce a collaborative multi-agent framework as a novel method of incorporating inference-time computation into AQG. This approach leverages multiple agents that iteratively refine generated question-answer pairs to better balance complexity and cognitive demand. We evaluate the generated questions on five meta-evaluation criteria: relevance, importance, clarity, difficulty matching, answerability, to assess the system's ability to control the required complexity and quality of the questions. Preliminary evaluations show that this collaborative multi-agent framework elevates the quality of generated educational content by fostering a more nuanced balance between cognitive challenge and clarity. These promising outcomes suggest that integrating collaborative multi-agent workflows can yield more controlled, pedagogically valuable content that can help advance automated educational content generation and adaptive learning environments.

\end{abstract}

\section{Introduction and related work}

\begin{figure}
\centering
\includegraphics[width=1\linewidth]{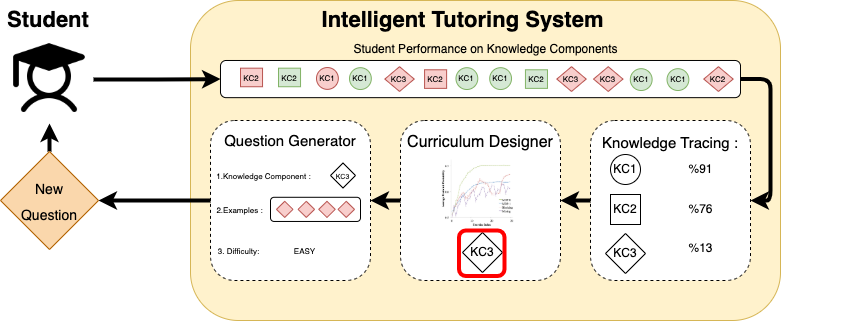}
\caption{Typical workflow of an ITS system.}
\label{fig:ITS}
\Description{Workflow of an ITS System.}
\end{figure}

Dynamically generating mathematics practice exercises and homework problems is a longstanding challenge for Intelligent Tutoring Systems (ITS) and educators \cite{heilman_good_2010}. Typically, educators and ITS either rely on existing problems or create new variations through minor modifications. Automatic Question Generation (AQG) aims to address this challenge but faced limited early success \cite{heilman_good_2010,zhang_review_2022}. Over time, diverse AQG methods have emerged, including the fine-tuning of pretrained language models for topic-controlled question generation \cite{li_novel_2025} \cite{vaswani_attention_2023}, often leveraging contrastive examples (examples of similar questions that differ in topic) to help the model distinguish and generate questions suited to specific concepts. More recently, prompt engineering guided by Bloom’s Taxonomy \cite{blobstein_angel_2024}, few-shot prompting with chain-of-thought reasoning \cite{scaria_automated_2024,blobstein_angel_2024}, and contextualized math word problem generation with equation constraints \cite{hwang_contextualized_2024,wang_math_2021} have further advanced the field. Modern ITS require robust AQG capabilities to dynamically provide personalized practice problems, effectively intervening in students' learning trajectories~\cite{piech_deep_2015}. \\
An ITS system can be conceptualized as managing a student's knowledge state. Figure \ref{fig:ITS} illustrates a typical ITS architecture where a knowledge-tracing module continuously monitors student comprehension across various Knowledge Components (KCs). At each interaction, the ITS selects a KC based on the student's knowledge state and using a Curriculum Designer module, commonly implemented via reinforcement learning (RL) models or search algorithms like expectiminimax \cite{piech_deep_2015}, to present an appropriate problem as an intervention. This adaptive process supports personalized curricula aimed at continuously enhancing student mastery of KCs. Consequently, ITS must be capable of dynamically generating practice problems for any KC, leveraging heuristics from example problems and incorporating difficulty assignment mechanisms. 
The AQG task can vary significantly based on system requirements. For instance, \cite{blobstein_angel_2024} utilizes Large Language Models (LLM) prompted explicitly by textbook chapters and related question sets, while other systems may rely solely on textbook chapters without additional prompts. Another common variation is concept-based generation, where AQG systems create questions derived directly from explicitly defined concepts or knowledge graphs. This paper approaches AQG from an ITS perspective, defining inputs as: 1) the KC name per Common Core State Standards definitions, 2) a set of example questions for that KC, and 3) the required difficulty level (easy, medium, or hard). The system outputs a generated question and its answer. \\
AQG aligns with broader natural language generation (NLG) challenges, particularly the lack of definitive ground truths and the subjective nature of human evaluations. Because of this, evaluating NLG has been a key area of recent research, giving rise to frameworks like G-Eval \cite{liu_g-eval_2023}, which leverages GPT-4 as a reference-free evaluator for tasks such as summarization, demonstrating improved alignment with human judgments. However, biases may persist in these LLM-based evaluation methods, as highlighted by \cite{koo_benchmarking_2024,wang_dhp_2025}. Task-specific initiatives like QGEval \cite{fu_qgeval_2024} reveal continued difficulties in ensuring strong alignment between automated metrics and human ratings in question generation, underscoring the broader challenges posed by subjectivity and the absence of definitive ground truth. Another major ongoing challenge in AQG is ensuring high-quality questions that are coherent, relevant, and accurately aligned with the intended difficulty level. \\
To address these issues, we explore recent advancements in  LLMs, particularly in Inference Time Computation (ITC) techniques \cite{snell_scaling_2024}. ITC generally refers to methods that extend or refine a model’s outputs during deployment without additional training, by orchestrating extra computational steps, retrieving context, or allowing models to interact with each other. Within this paradigm, we focus on multi-agent collaborative systems, also known as agentic workflows \cite{wu_autogen_2023}, which improve LLM performance by facilitating interactions between multiple agents. This allows them to debate, complement each other’s strengths, and incorporate diverse perspectives. Of all the agent collaboration techniques, debate emerges as a particularly effective approach, inspired by the concept of the Society of Mind \cite{minsky_society_1988} to harness collective knowledge. Studies show that multi-agent debate can enhance factuality and reasoning \cite{du_improving_2023}, foster divergent thinking \cite{liang_encouraging_2024}, and even achieve state-of-the-art performance in mathematical reasoning \cite{wu_autogen_2023}. Other lines of work focus on mechanisms such as hierarchical or roleplaying-based collaboration \cite{zhang_review_2022,chen_oscars_2025} and competitive settings \cite{zeng_autodefense_2024}, while recent frameworks like AutoGen \cite{wu_autogen_2023}, Camel \cite{li_camel_2023}, and MetaGPT \cite{hong_metagpt_2024} facilitate a variety of multi-agent collaboration structures. By enabling agents to engage in structured interactions, these workflows enhance factual accuracy, reasoning capabilities, and the overall quality of generated questions. \\
In this paper, we investigate the potential of integrating ITC techniques through collaborative multi-agent frameworks into AQG. We evaluate the outputs based on clarity, relevance, importance, answerability, and difficulty matching. Our contributions are three-fold. First, we propose two novel collaborative frameworks specifically designed for AQG tasks. Second, we develop a self-curation method for AQG guided by cognitive demand criteria from Bloom's taxonomy. Third, we introduce an automated evaluation framework to assess automatically generated math questions, considering clarity, relevance, importance, answerability, and difficulty control.
Our research explores two key questions: \\ (1) Does difficulty matching depend on the proposed difficulty? \\
(2) Do agentic workflows improve question generation compared to baseline models?
\section{Method and Experiment Setup}
We propose two multi-agent systems that generate a question and answer based on a given difficulty level, set of examples, and the KC name for use in the ITS. In this section, we will explain the data, agents, multi-agent workflows, prompting strategies, curation method, and evaluation metrics of the system.
\subsection{Data}
Our experiments utilize the Problem Bodies~\footnote{\url{https://sites.google.com/site/assistmentsdata/home/assistments-problems}} dataset. The dataset is an extension of the ASSISTments dataset, a widely used benchmark from ITS. Problem Bodies contains middle-school math questions accompanied by empirical student performance metrics. Each question in this dataset includes a ``percent correct" attribute, representing the percentage of students who answered correctly. This allows us to categorize the questions into easy, medium, and hard difficulty levels, providing an empirically grounded measure of difficulty aligned with real-world student performance.

\subsection{Agents} In our experiments, we use the following agents/roles:\\
\textbf{Teacher:} Generates math questions and answers for a specific KC at a set difficulty level. It returns concise responses while adjusting based on conversation context.\\ 
\textbf{Generic Critic:} Provides objective, high-level feedback on clarity, relevance, and difficulty alignment without introducing new content. It references the overall discussion for consistent evaluation.\\
\textbf{Consensus CEO:} Serves as the final decision-maker by reviewing conversation history and selecting the best question-answer pair from multiple agents. If a consensus exists, it reports that; otherwise, it chooses the option aligning best with the KC, difficulty requirements, and sample solutions.\\
\textbf{Versatile Agent:} Dynamically participates in collaborative discussions, making one of three decisions based on chat history: (1) generate a new Q\&A pair, (2) revise an existing pair, or (3) endorse a peer’s pair while providing constructive feedback. This agent uses prior messages and examples to ensure its contributions align with the KC, intended difficulty, and the collective goal of reaching consensus.

\subsection{Workflows and Experimental Setup}
\begin{figure}
    \centering
    \includegraphics[width=0.8\linewidth]{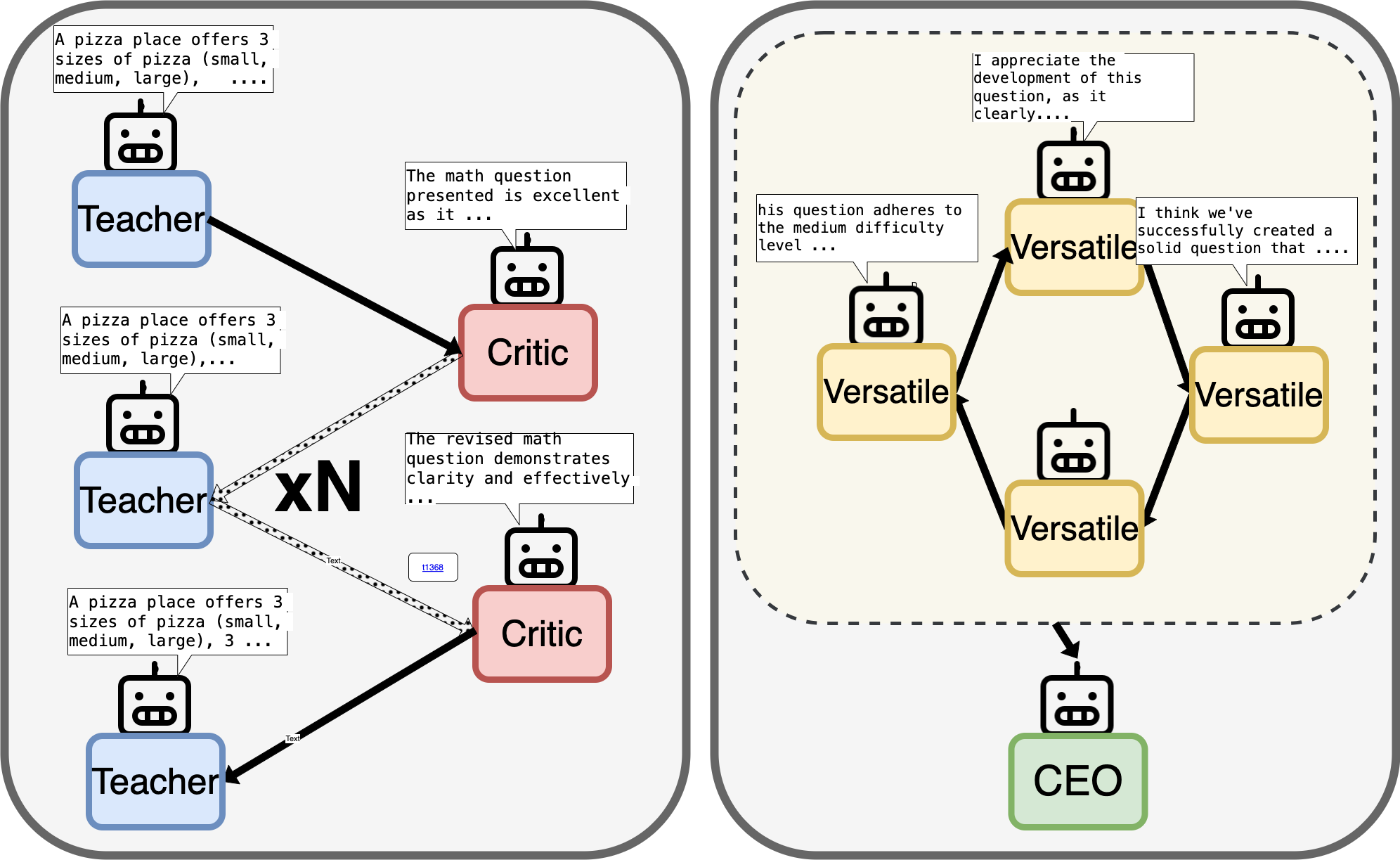}
    \caption{Agentic workflows. Left: Teacher Critic Cycle. Right: Collective Consensus.}\label{fig:agentic_workflows}
    \Description{See Section 2.3 for Details.}
\end{figure}We systematically evaluated four distinct approaches, including two Baseline scenarios and two agentic workflows (see Figure \ref{fig:agentic_workflows}) designed to generate educational math question-answer pairs, each with varying complexity and collaborative structures: \\
\textbf{Baseline Teacher Zero-Shot:}
This workflow uses a single teacher agent using a zero-shot prompt approach to generate self-contained math questions and answers without relying on prior examples or conversation history.\\
\textbf{Baseline Teacher Few-Shot:}
In this variant, the teacher agent is guided by a few example question-answer pairs, aligning it more closely with other few-shot-based workflows.\\
\textbf{Teacher-Critic Cycle (TCC):}
TCC involves two agents: a teacher generates an initial math question and answer, followed by iterative critique from a generic critic agent. The critic assesses clarity, relevance, difficulty, and pedagogical appropriateness, with the teacher refining the content across multiple feedback rounds. Experimentally, we varied the number of interaction rounds between two to five, and evaluated the effectiveness of two prompt engineering techniques: Auto Chain-of-Thought (AutoCoT) and explicit Solution Generation, by enabling and disabling them. In Solution Generation, the teacher (or versatile agent) tries to output only the final answer to a math problem, while in AutoCoT, the teacher (or versatile agent) uses a step-by-step reasoning process before arriving at the final answer. \\
\textbf{Collective Consensus (CC):}
CC is a multi-agent workflow involving a collaborative conversation to reach consensus on math question-answer pairs. Initially, one versatile agent generates a question-answer pair, after which two to four versatile agents sequentially contribute by either creating a new question, revising an existing question with feedback, or explicitly agreeing with feedback. Decoding parameters, including sampling seed and temperature, are randomized per agent to encourage diverse perspectives. Following iterative discussion (ranging from two to five rounds), a consensus CEO agent reviews the conversation. If consensus is achieved, the CEO selects the agreed-upon final pair; otherwise, it chooses the best candidate based on collective judgment and educational alignment. As with TCC, we explored the impact of AutoCoT and explicit Solution Generation, evaluating their roles both enabled and disabled.
Across all workflows, we tested three prompting strategies which we detail in Section~\ref{Difficulty Prompting Strategies}: Empirical Difficulty Prompting, Empirical Prompting, and Simple Prompting. These strategies help assess whether providing explicit difficulty-level context enhances question quality relative to random or difficulty-specific examples. 
\subsection{Difficulty Prompting Strategies}
\label{Difficulty Prompting Strategies}
\textbf{Empirical:} All but the baseline use a few-shot approach. Math questions are labeled “easy,” “medium,” or “hard” using real student performance data (higher ``percent correct" indicates easier). These labeled examples guide the model in generating questions matching the requested difficulty.\\
\textbf{Prompting Empirical:} This variant also relies on empirical difficulty data but only presents examples matching the requested difficulty. For instance, if the model is asked to produce an “easy” question, only empirically identified easy examples are shown. \\
\textbf{Prompting Simple (Random Examples):} Here, the model receives randomly selected examples from all difficulty tiers. It must rely solely on the requested difficulty (“easy,” “medium,” or “hard”) without explicit guidance from example labels, placing more responsibility on general instructions.
\subsection{Self-Curation}
We design a \textbf{Bloom Agent} that assigns each output question a Bloom evaluation score (1–5) based on its cognitive demands, ensuring alignment with the three Bloom tiers—lower (Remembering, Understanding), middle (Applying, Analyzing), and upper (Evaluating, Creating).  Because generating many candidate questions is computationally cheap, we aggressively discard those that fail the expected cognitive challenge, retaining only those demonstrating strong Bloom alignment. We also compare this curation method against Random Curation (RC) of our methods for comparison. Unlike conventional curation, which may rely on syntactic features, our approach prioritizes pedagogical rigor over mere textual similarity, ensuring that even a surplus of suboptimal questions yields a subset with meaningful cognitive depth. 
\subsection{Evaluation Metrics}
To evaluate the generated questions, we used prompt engineering to quantify the quality of each instance based on the following criteria: \\
\textbf{Relevance:}
Measures how well the generated questions align with example questions and the targeted KC. Scored on a scale of 1 (not relevant) to 5 (highly relevant). \\
\textbf{Importance:}
Assesses whether the question emphasizes key conceptual components of the targeted KC. Rated from 1 (least important) to 5 (essential). \\
\textbf{Clarity:}
Evaluates structural coherence and linguistic precision of a question to ensure it is easily understood by middle-school students. Scored from 1 (unclear) to 5 (exceptionally clear). \\
\textbf{Difficulty Matching:}
Determines how well a question’s complexity aligns with the specified difficulty level (easy, medium, hard). Cognitive difficulty ranges from basic factual recall (easy) to moderate conceptual reasoning (medium) and advanced analytical skills (hard). Scored from 1 (no alignment) to 5 (perfect alignment). \\
\textbf{Answerability:}
Evaluates whether a middle-school student can reasonably understand and respond to the question based on the provided information. Rated from 1 (unanswerable) to 5 (clearly answerable). \\
The prompts for the agents and the evaluation module can be found in our repository~\footnote{\url{https://github.com/aminsmd/QA_GEN}}

\section{Results and Discussion}
\begin{figure*}
    \centering
    \includegraphics[width=0.7\linewidth]{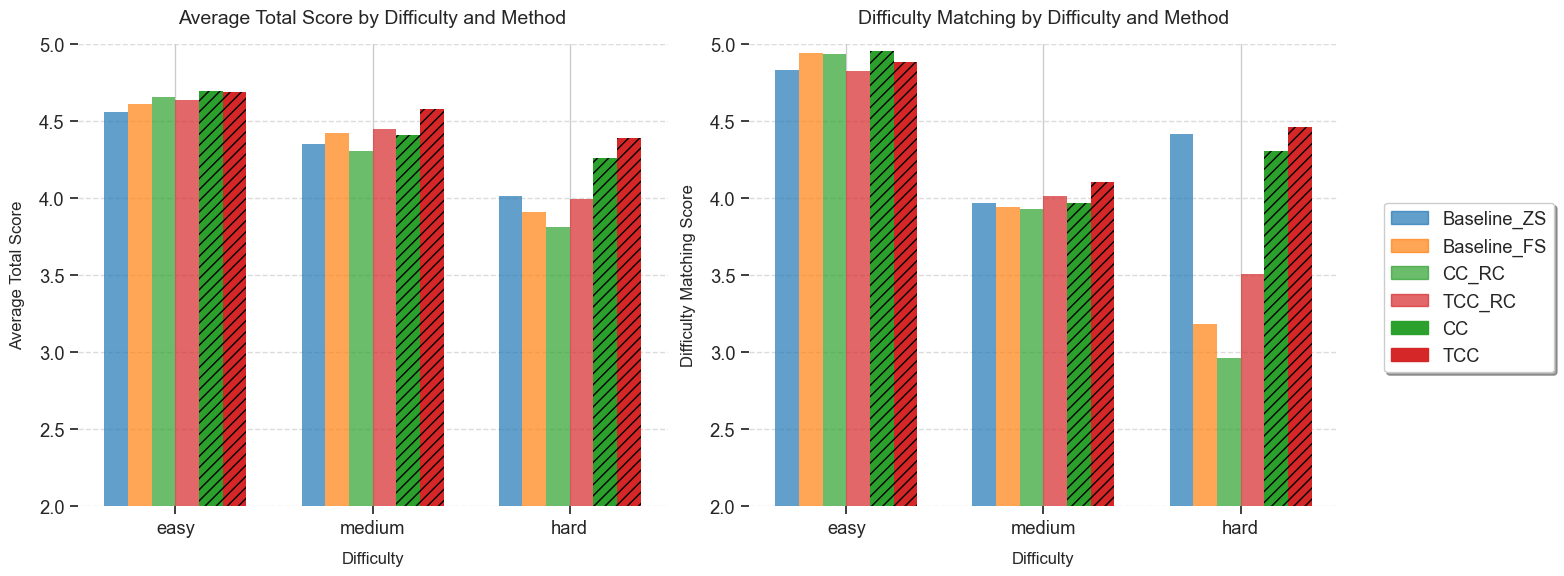}
    \Description{Bar graph showing average score and difficulty matching by difficulty and methods.}
    \caption{Average score and difficulty matching by difficulty and methods.}
    \label{fig:res_bargraph}
\end{figure*}

\begin{table*}
\centering
\small  
\renewcommand{\arraystretch}{0.7} 
\setlength{\tabcolsep}{4pt} 
\caption{Evaluation results of different question generation workflows.}
\label{tab:results}
\begin{tabular}{lcccccc}
\toprule
\textbf{Method} & \textbf{Clarity} & \textbf{Relevance} & \textbf{Importance} & \textbf{Difficulty Matching} & \textbf{Answerability} & \textbf{Avg. Score} \\
\midrule
\textbf{Baseline\_ZS}  & 3.66 & 4.61 & 4.67 & 4.41 & 4.65 & 4.40 \\
\textbf{Baseline\_FS}  & 3.70 & 4.93 & \textbf{4.73} & 4.02 & 4.71 & 4.42 \\
\textbf{CC\_RC}        & 3.50 & \textbf{4.95} & 4.71 & 3.94 & 4.61 & 4.34 \\
\textbf{TCC-RC}        & \textbf{3.72} & 4.90 & \textbf{4.73} & 4.11 & 4.79 & 4.45 \\
\textbf{CC}            & 3.60 & \textbf{4.99} & \textbf{4.76} & \textbf{4.96} & \textbf{4.94} & \textbf{4.65} \\
\textbf{TCC}           & \textbf{3.75} & 4.92 & 4.70 & \textbf{4.88} & \textbf{4.94} & \textbf{4.64} \\
\bottomrule
\end{tabular}
\end{table*}

Table \ref{tab:results} shows the evaluation results of our methods, TCC and CC, compared to both the baseline and non-curated agentic methods. While TCC and CC outperform both the baselines and their non-curated counterparts across all evaluation metrics, the improvements are incremental rather than drastic. The greatest gains are observed in Difficulty Matching and Relevance, which suggests that incorporating iterative critique and collective refinement enhances alignment with the intended cognitive challenge and pedagogical quality. These findings directly address Research Question 2 (RQ2). Additionally, the fact that non-curated agentic methods underperform the baseline suggests that agent-based generation alone is insufficient without structured selection or refinement. Agent responses exhibit greater variability, sometimes failing to meet standards of clarity, difficulty alignment, or answerability. This highlights the importance of iterative curation mechanisms to ensure consistency and reliability in multi-agent AQG. \\
Figure \ref{fig:res_bargraph} presents the Difficulty Matching and Average Score across different difficulty levels (easy, medium, hard) for each method. The trends observed align with the overall evaluation results. Our curated methods (TCC and CC) outperform the baselines and non-curated methods, though the improvements remain incremental. A general pattern emerges where both Difficulty Matching and Average Score decrease as question difficulty increases. This finding addresses Research Question 1 (RQ1). The observed trend suggests that difficulty matching is indeed more challenging for harder questions, indicating that LLM-based generation struggles to maintain appropriate cognitive complexity as difficulty increases. Notably, the baseline models and non-curated agentic methods perform significantly worse on hard questions. This reinforces the need for structured prompting and iterative refinement to produce well-calibrated high-difficulty questions. \\
Unexpectedly, the Baseline Zero-Shot method achieves higher Difficulty Matching scores than both the Baseline Few-Shot model and several agentic methods (including curated and non-curated variants), which suggests that few-shot examples may introduce biases or inconsistencies in aligning questions with intended difficulty. These findings underscore the intricate interplay between prompting strategies and difficulty alignment, prompting further study of how example-based prompts shape model outputs.
A key question involves how much inference computation is optimal. Since these are iterative ITC methods, merely increasing the number of rounds or agents does not guarantee performance gains and can even yield diminishing returns (Figure \ref{fig:combined_plot}). Our limited settings suggest that a more comprehensive parameter search is needed to pinpoint where additional computation yields real benefits.\\ 
Furthermore, our study finds that the effectiveness of prompting strategies for enabling the in-context learning ability of LLMs is minimal. As shown in Table \ref{tab:difficulty_method}, different few-shot prompting techniques had little to no impact on performance. This suggests that current few-shot learning strategies may be suboptimal for the AQG task. Future research could explore alternative prompt engineering and adaptive few-shot learning techniques to better leverage in-context learning for question generation. \\
Finally, a critical area for future work is the quality of evaluation itself. All our findings rely on automated evaluations conducted by GPT-4 and based on state-of-the-art NLG evaluation frameworks \cite{liu_g-eval_2023,fu_qgeval_2024}, which raises concerns about potential biases and limitations in LLM-based self-evaluation. In particular, we observed a ceiling effect (Figure \ref{fig:histogram}) in the resulting evaluation scores similar to what the G-Eval authors reported. This effect can mask finer-grained differences in system performance and limit the discriminative power of LLM-based evaluations. If the evaluation system exhibits misalignment with human judgment, all derived insights could be affected. To address this, we plan to collect human evaluation data and fine-tune our automated evaluation module to align more closely with human evaluators and increase the trustworthiness of the assessment process.\\
Overall, while our study demonstrates that the success of difficulty matching is dependent on the proposed difficulty level (RQ1) and that agentic workflows can improve question generation compared to the baseline (RQ2), these improvements come with important caveats. Optimization of inference computation, refinement of few-shot learning strategies, and enhancing evaluation reliability are key directions for future research to advance AI-driven educational content generation.

%
\bibliographystyle{abbrv}
\bibliography{references.bib}  
%
\appendix
\section{Additional Figures and Results}

\begin{figure*}[h]
\begin{subfigure}{0.32\textwidth}
  \centering
  \includegraphics[width=\linewidth]{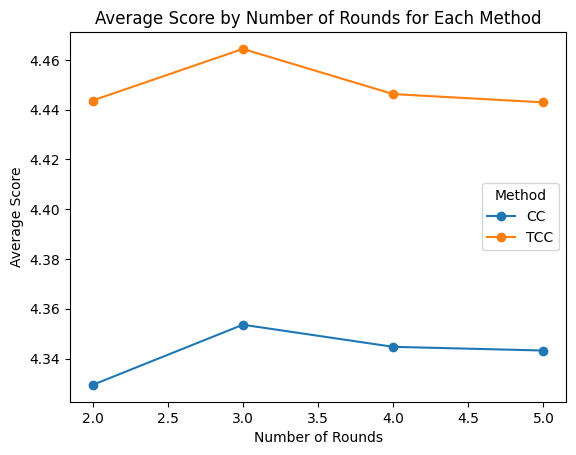}
    \caption{}
    \label{fig:rounds}
    \Description{Average Score by Number of Rounds for Each Method.}
\end{subfigure}\hfill
\begin{subfigure}{0.32\textwidth}
  \centering
  \includegraphics[width=\linewidth]{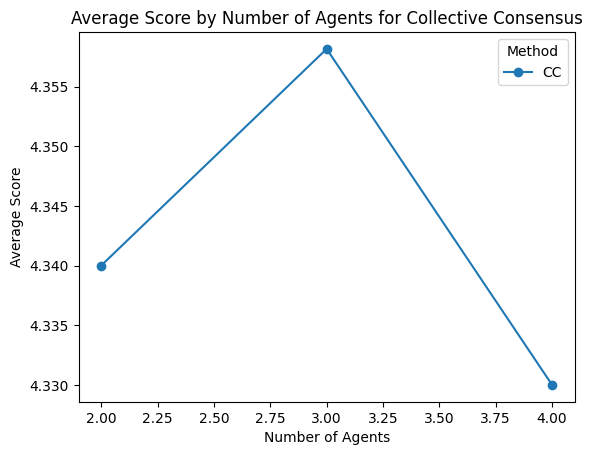}
    \caption{}
    \label{fig:agents}
    \Description{Average Score by Number of Agents for Collective Consensus.}
\end{subfigure}\hfill
\begin{subfigure}{0.32\textwidth}
  \centering
  \includegraphics[width=\linewidth]{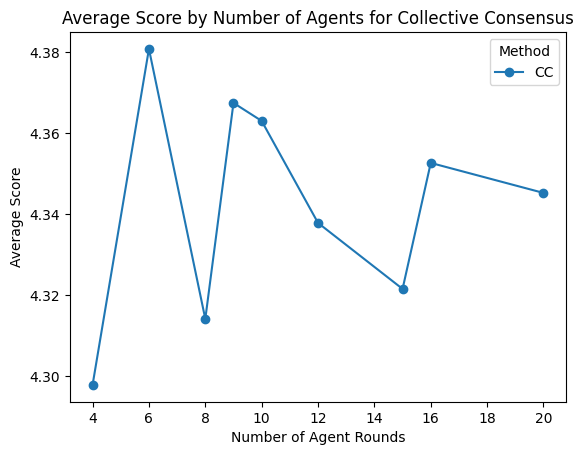}
    \caption{}
    \label{fig:n_agent_round}
    \Description{Average Score by Number of Agent Rounds for Collective Consensus.}
\end{subfigure}\hfill
\caption{Average score plotted against: (a) number of agentic discussion rounds for both TCC and CC methods, (b) number of agents in CC workflow, and (c) number agentic rounds (number of agents multiplied by number of rounds).}
\label{fig:combined_plot}
\end{figure*}

\begin{figure*}[h]
    \centering
    \includegraphics[width=0.7\linewidth]{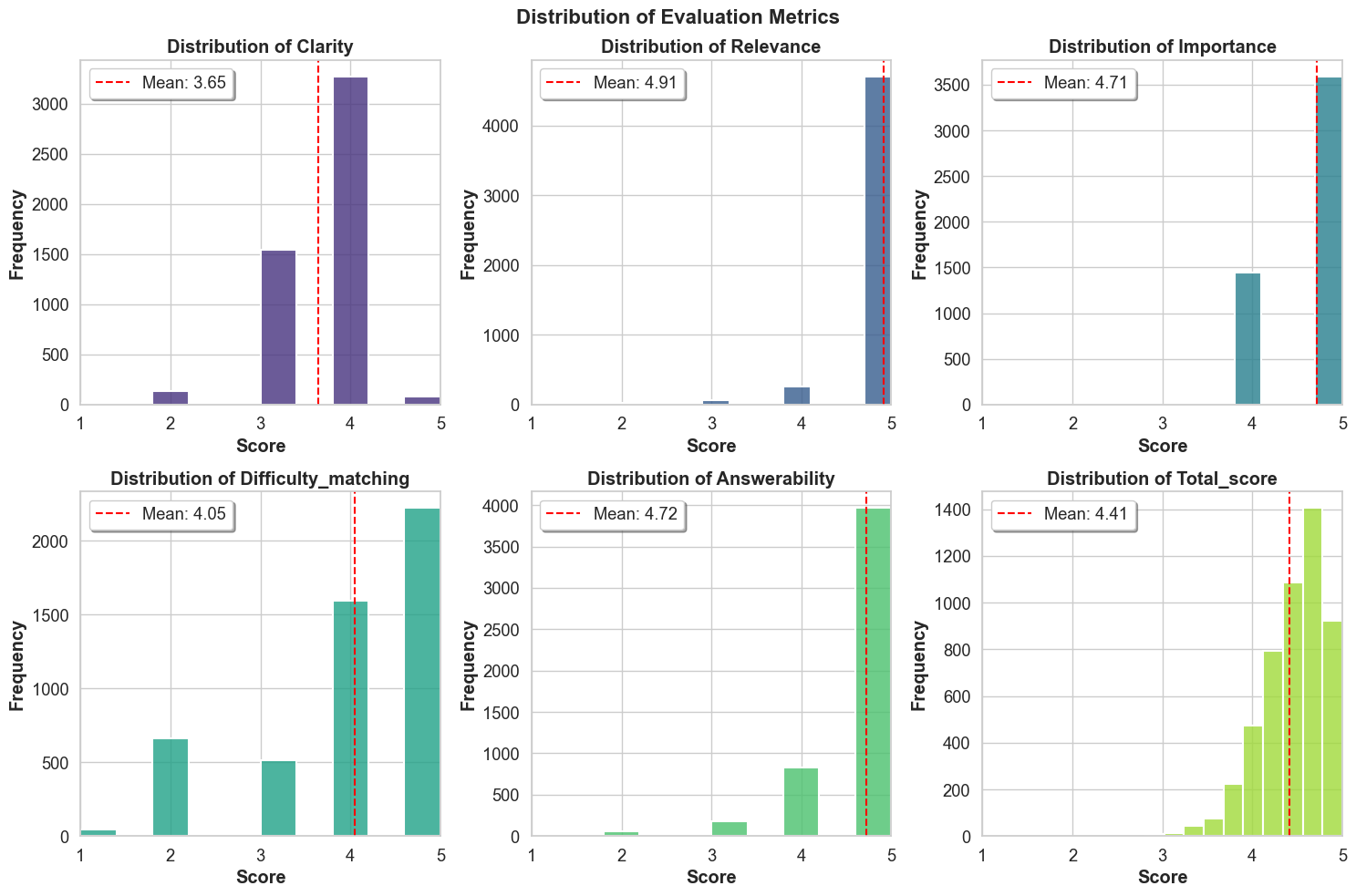}
    \caption{Histogram of evaluation metrics.}
    \label{fig:histogram}
    \Description{Distribution of evaluation metrics.}
\end{figure*}

\begin{table*}
\centering
\caption{Difficulty prompting strategies results.}
\label{tab:difficulty_method}
\begin{tabular}{llcc}
\toprule
\textbf{Method} & \textbf{Difficulty Prompting Strategies} & \textbf{Difficulty Matching} & \textbf{Avg. Score} \\
\midrule
CC & empirical & \textbf{4.71} & 4.60 \\
CC & prompting empirical & 4.68 & 4.64 \\
CC & prompting simple & \textbf{4.70} & 4.61 \\
TCC & empirical & 4.66 & \textbf{4.66} \\
TCC & prompting empirical & 4.61 & 4.64 \\
TCC & prompting simple & 4.67 & \textbf{4.65} \\
\bottomrule
\end{tabular}
\end{table*}

\balancecolumns
\end{document}